\documentclass[%
 reprint,
showpacs,
preprintnumbers,
 amsmath,amssymb,
 aps,
]{revtex4-1}

\usepackage{graphicx}
\usepackage{dcolumn}
\usepackage{bm}
\usepackage[T1]{fontenc}
\usepackage{textcomp}

\usepackage{amsfonts}
\usepackage{upgreek}     
\usepackage[T1]{fontenc} 
\usepackage{ae,aecompl}  

\newcommand{\micron}{~\ensuremath{\upmu\text{m}}}

\begin{document}

\title{An electrostatic elliptical mirror for neutral polar molecules}
\author{A. Isabel Gonz\'alez Fl\'orez}
\author{Samuel A. Meek}
\author{Henrik Haak}
\author{Horst Conrad}
\author{Gabriele Santambrogio}
\email{gabriele.santambrogio@fhi-berlin.mpg.de}
\author{Gerard Meijer}
\affiliation{Fritz-Haber-Institut der Max-Planck-Gesellschaft,
Faradayweg 4-6, 14195 Berlin, Germany}
\date{\today}

\begin{abstract}
Focusing optics for neutral molecules finds application in shaping and
steering molecular beams. Here we present an electrostatic elliptical 
mirror for polar molecules consisting of an array of microstructured 
gold electrodes deposited on a glass substrate. Alternating positive 
and negative voltages applied to the electrodes create a repulsive 
potential for molecules in low-field-seeking states. The equipotential 
lines are parallel to the substrate surface, which is bent in an elliptical 
shape. The mirror is characterized by focusing a beam of metastable 
CO molecules and the results are compared to the outcome of trajectory 
simulations.
\end{abstract}

\maketitle

\section{Introduction}
Miniaturization of electric-field structures enables the creation of large 
field gradients with moderate voltages, that is, large forces and tight 
potential wells for neutral polar molecules. Moreover, present day 
microstructuring technology offers the possibility to integrate multiple 
tools and devices onto a compact surface area. Recently, guiding, deceleration and 
trapping of polar molecules on a 2~cm$^2$ area on a chip have been 
demonstrated \cite{Meek:2008p153003,Meek:2009p055024,Meek:2009p1699}. 
In these experiments molecules have been loaded into microscopic 
traps, located a few tens of micrometers above the surface of the chip, 
directly from a pulsed supersonic molecular beam. The density of the 
molecules trapped on the chip is determined by the density of the 
molecular beam near the entrance of the chip. To increase the density 
of the molecules on the chip, the chip could be moved closer to the 
molecular beam source. Collisions in the expansion region of the molecular 
beam and the need for differential pumping stages and skimmers in the 
beam machine limit the minimum distance of the chip from the source, 
however. It would be advantageous, therefore, to position a focusing 
element between the source and the molecule chip to increase the density 
of molecules on the chip. This focusing element would have the added 
advantage of being state-selective, i.e., of purifying the molecular beam 
and of selectively focusing molecules in the desired quantum state onto 
the molecule chip.

Electrostatic quadrupoles \cite{Gordon:1954p282,Bennewitz1955p6} and
hexapoles \cite{Kramer:1965p767} have long been used for focusing and
quantum state selection of beams of polar molecules. The electric
field inside these multipoles is in a good approximation cylindrically
symmetric and they thereby act as positive (negative) spherical lenses
for molecules in low-field-seeking (high-field-seeking)
states. Retroreflection of beams of neutral polar molecules with
electric and magnetic mirrors has recently also been reported
\cite{Schulz2004p020406,Metsala:2008p053018}. Even though designs for
such mirrors had already been discussed since the 1950s
\cite{White:1959p596}, their experimental demonstration had to await
the advent of sufficiently intense beams of slow polar molecules
\cite{Meerakker:2008p595}. 

Here we present an electrostatic elliptical mirror for neutral polar
molecules in low-field-seeking states. Its operation is based on the
same principle as that of multipole focusers, but its focusing
properties are more ideally suited to couple a molecular beam into the
elongated, cylindrical traps on a molecule chip. Its simple
implementation makes it a versatile tool that will most likely find
application in other experimental setups as
well.

\section{An electrostatic mirror}
A planar electrostatic mirror consisting of an array of parallel and
equidistant electrodes on a surface to which alternating voltages,
$\pm V_0$, are applied was first discussed by Wark and Opat
\cite{Wark1992p4229}. If the electrodes are sufficiently long in, say,
the $x$-direction, end-effects can be ignored and the problem of
calculating the electric potential produced by these electrodes can be
performed in two dimensions. With this simplification, the electric
potential in the ($y$,$z$)-plane can be written as  
\begin{equation}\label{eq:elpot}
V(y,z)=V_0\sum_{n=0}^\infty \bigg[A_{2n+1}\cos\bigg(\frac{2\pi(2n+1) z
  }{\ell}\bigg) 
e^{-\frac{2\pi (2n+1)y}{\ell}}\bigg],
\end{equation}
where the $y$-direction is chosen to be perpendicular to the plane of
the electrodes, the $A_{2n+1}$ are dimensionless parameters that
depend on the specific geometry of the array, and $\ell$ is the period
of the array. By taking only the first two terms of
Equation~\eqref{eq:elpot} into account, the resulting magnitude of the
electric field is 
\begin{multline}\label{eq:elfield}
E(y,z)=A_1V_0\frac{2\pi}{\ell} e^{-\frac{2\pi y}{\ell}} \\
\sqrt{1 + \frac{6A_3}{A_1} \, e^{-\frac{4\pi y}{\ell}}
\cos\bigg(\frac{4 \pi z}{\ell}\bigg)+
\bigg(\frac{3A_3}{A_1}\bigg)^2 \, e^{-\frac{8\pi y}{\ell}}}.
\end{multline}
The magnitude of the electric field thus exponentially decays with
distance from the surface while being periodic along the surface. The
dependence of the  magnitude of the electric field on $z$ vanishes for
$y\gg\ell/ 4\pi$. In general, the force that neutral polar molecules
experience in an electric field is given by the negative gradient of
the Stark energy of the quantum level that they are in. This force can
be expressed as the product of the gradient of the electric field
strength with an effective dipole moment,
$\mu_\textrm{eff}$. The latter is negative for molecules in
low-field-seeking quantum states and these molecules are therefore
repelled from the surface. If the component of their kinetic energy
towards the surface is low enough, they will be turned around before
ever getting close to the surface and they will only experience a
$z$-independent potential. In the present experiments the molecules
have a high forward velocity but they approach the surface at grazing
incidence. As a result, the molecules sample many periods of the
electrode array while approaching the surface, so the effective
electric field they experience is averaged over many periods. The
corrugation of the electric field strength close to the surface can
then be ignored, and the second term under the square-root in
Equation~\eqref{eq:elfield} averages to zero. For the experiments
reported here we have reshaped a planar microstructured electrostatic
mirror into an elliptical one in the ($y$,$z$)-plane; in the
($x$,$z$)-plane it has been kept flat. By placing the (point) source
of a divergent molecular beam at one focus of the elliptical mirror,
molecules can be reflected to the other focus, creating a line-focus
along the $x$-direction in the present setup.
Elliptical mirrors have been used to focus beam of particles
before~\cite{Fladischer2010p033018,Yamamura2010p193} but this is---to
the best of our knowledge---the first demonstration of an
electrostatic elliptical mirror.

\section{Experimental setup}
\begin{figure}
\centering
\includegraphics[width=.45\textwidth]{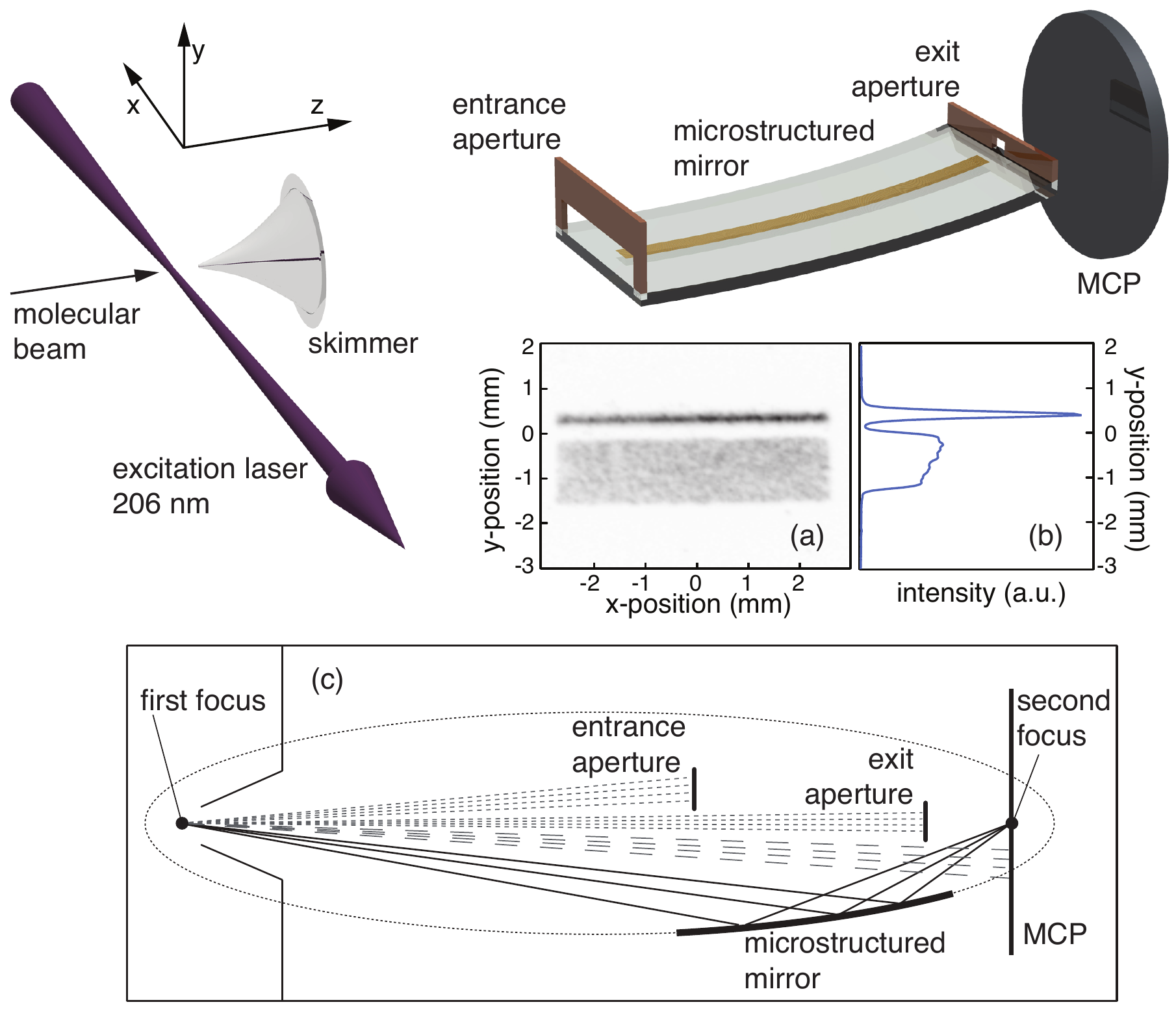}
\caption{Schematic of the experimental setup (top). Laser-prepared
  metastable CO molecules reach the microstructured elliptical mirror
  after passing through a 1~mm skimmer and are detected further downstream
  by an imaging MCP with phosphor screen. The time-integrated density of
  metastable CO molecules is recorded by a CCD camera (a) and integrated 
  along the $x$-axis (b). Panel (c) at the bottom shows how the
  different parts of the molecular beam contribute to the recorded
  signal in the ideal case of a point source (figure not drawn to
  scale).\label{fig:Setup}}
\end{figure}

A scheme of the experimental setup is shown in Figure
\ref{fig:Setup}. A pulsed molecular beam is generated by expanding
20\% CO seeded in krypton or neon through the nozzle of a pulsed
supersonic valve (Jordan ToF Products, Inc.). The CO molecules are
excited to the upper $\Lambda$-doublet component of the $a^3\Pi_1$,
$v'=0$, $J'=1$ ($J'=2$) level directly from the $N"=1$ ($N"=2$)
rotational level of the electronic and vibrational ground state, 
using narrow-band pulsed laser radiation around 206~nm (1~mJ in a
5~ns pulse with a bandwidth of about 150~MHz). After passing through a 
skimmer with a 1~mm diameter opening, the metastable CO molecules,
which have a radiative 
lifetime of about 2.6~ms \cite{Gilijamse:2007p1477}, reach the
elliptical mirror. This microstructured mirror consists of an array of
1254 gold electrodes, each of which are 4~mm long, 10\micron\ wide,
and approximately 100~nm high. The electrodes are deposited with a
40\micron\ center-to-center spacing onto a 1~mm thick glass substrate,
creating an active reflective area of 4~mm by 50~mm (Micro Resist
Technology GmbH) \cite{Meek:2009p055024}. The parameters $A_1$, $A_3$ and
$\ell$ in Equation~\eqref{eq:elfield} for this geometry are 1.196,
0.2478, and 80\micron, respectively. The glass
substrate is clamped between two bronze pieces that have been machined
such that they bend the glass into an elliptical segment. The ellipse
has a major axis of 245.4~mm and a minor axis of 7.1~mm. The
elliptical mirror with its holder is mounted on a manipulator that
allows the mirror to be translated parallel to either one of the axes
of the ellipse. It also allows the angle $\chi$ between the long axis
of the ellipse and the molecular beam axis to be accurately
adjusted. The movements are actuated by four vacuum compatible piezo
motors (New Focus, Model 8354). At the entrance of the mirror there is
a 3.5~mm high entrance slit; near the exit there is another aperture,
4~mm wide and 1.2~mm high. Behind the mirror, the spatial distribution
of the metastable CO molecules in the ($x$,$y$)-plane is recorded with
a 2'' imaging micro-channel plate (MCP) detector with phosphor screen
and imaged by a fast CCD camera. The detector can be gated by rapidly
switching the voltage applied to the entrance surface of the MCP. In
most measurements, the signal is integrated over the full arrival time
distribution of the molecules in the pulse. In a typical arrangement,
the mirror is positioned such that one focus of the ellipse is near
the laser excitation point, the other focus is near the MCP detector
and the angle $\chi$ is around zero degrees. Panel (a) in
Figure~\ref{fig:Setup} shows a gray-scale image of the intensity of
the reflected (upper horizontal stripe) and non-reflected (bottom
rectangle) molecular beam as recorded by the CCD camera. Next to it,
in panel (b), the signal is shown as a function of the $y$-coordinate
only, i.e., after integrating the signal along the $x$-axis. The
narrow peak corresponds to the signal of the reflected molecules, and
the broad plateau to that of the direct beam. Although we could
obviously have made the apertures on the elliptical mirror smaller,
thereby avoiding transmission of the direct beam all together, having
the direct beam is useful for alignment purposes as well as to have an
internal calibration for the amount of reflected molecules.  

Before installing the elliptical mirror, we characterized the free
molecular beam, 
using the gated two-dimensional imaging detector. For the beam of CO 
seeded in neon a Gaussian velocity distribution with a mean velocity
of 1000~m/s and with a full width at half maximum (FWHM) of 100~m/s is
found. The transverse velocity distribution is also found to be
Gaussian and the angular distribution has a FWHM of
33~mrad. When seeded in krypton the mean velocity of the CO molecules
is reduced to 560~m/s, their FWHM velocity spread is about 60~m/s and
the angular distribution is broadened to a FWHM of 60~mrad. 

\section{Measurements and simulations}
After placing the mirror in the molecular beam machine, we
systematically optimized its position for optimum focusing of CO
molecules with a mean velocity of 1000~m/s with $V_0$=150~V applied to
the electrodes. The best focusing is obtained when the center of the
ellipse is 1.2~mm above the molecular beam axis. The narrowest waist
of the reflected beam is achieved when the first focus of the ellipse
is placed close to the nozzle while the MCP is a few millimeters in
front of the second focus. This effect, namely that the tightest waist
is not at the second focus of the ellipse, is also seen in the
trajectory simulations and is due to aberrations induced by the fact
that the molecular source is not a point source. 

\begin{figure}
\centering
\includegraphics[width=0.49\textwidth]{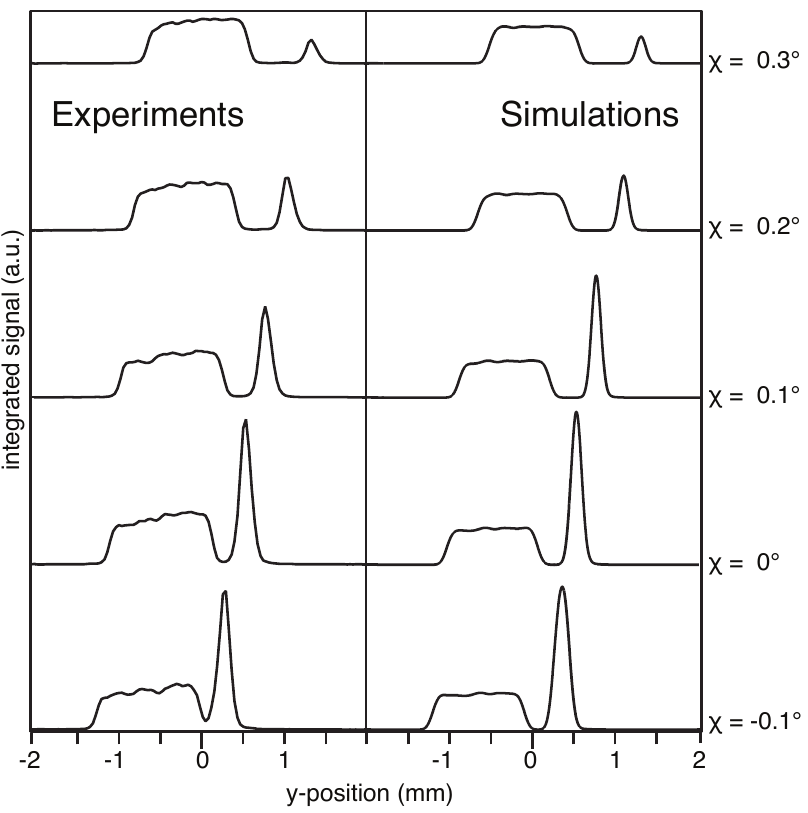}
\caption{Measured signal on the two-dimensional detector as a function
  of vertical position, i.e., integrated along the $x$-axis, for
  different tilt angles $\chi$ (left), together with the corresponding
  simulations (right). For these experiments a beam of metastable CO
  molecules with a mean velocity of 1000~m/s is used. The CO molecules
  are in the $J'=1$ level and voltages of $\pm$ 150~V are applied to
  the electrodes. \label{fig:tilt}}
\end{figure}

On the left-hand side of Figure~\ref{fig:tilt}, the measured signal
intensity is shown as a function of the $y$-position, i.e., after
integrating the two-dimensional image along the $x$-axis, for
different values of the tilt angle $\chi$. The largest intensity gain
of the reflected beam relative to the direct beam under these
conditions is about a factor 3, found at a tilt angle $\chi$ close to
zero. For more negative angles the intensity of the reflected beam
broadens and eventually merges with the direct beam (measurements not
shown). At more positive angles the intensity of the reflected beam
gradually decreases as the velocity component perpendicular to the
mirror surface increases and the molecules crashes into the
substrate. 

Trajectory calculations have been performed to simulate the recorded
two-dimensional images and the height profiles. For these
calculations, the known experimental geometry, the measured velocity
distribution of the free beam and the strength of the electric field
$E$ above the elliptical mirror as given by
Equation~\eqref{eq:elfield} are taken as input. The effective dipole
moment, $\mu_\textrm{eff}$, of the metastable CO molecules in
the selected rotational levels is the negative derivative of the Stark
energy of these levels with respect to the strength of the electric
field and can be written as 
\begin{equation}\label{eq:mueff}
\mu_\text{eff}(E) = - \cfrac{\Bigg(\cfrac{\mu\, \Omega
    M}{J(J+1)}\Bigg)}{\sqrt{ 1 + \Bigg(\cfrac{\Lambda}{2}\cfrac{J (J+1)}{E\,\mu\,\Omega M}\Bigg)^2}}.
\end{equation}
In this expression, $\Lambda$ is the magnitude of the
$\Lambda$-doublet splitting (394~MHz for the $J'=1$ and 1151~MHz for
the $J'=2$ level~\cite{Wicke:1972p5758}), $\Omega$M is the product of
the projection of the electronic angular momentum on the internuclear
axis with the projection of the total angular momentum $\vec{J}$ on an
external axis and $\mu$ is the magnitude of the body-fixed electric
dipole moment in the $a^3\Pi$ state (1.37~Debye~\cite{Wicke:1972p5758}).

Simulated height profiles for different values of the tilt angle are
shown on the right hand side of Figure~\ref{fig:tilt}. The main
features observed in the experiment are reproduced by the trajectory
simulations. The calculated distributions of reflected molecules have
a minimum width of about 200\micron\ (FWHM), implying a reduction of
about a factor five relative to the size of the source. The fraction of
molecules that approach the surface of the elliptical mirror at
$\chi=0^{\circ}$ closer than the distance $\ell/4\pi$ ($\approx 6\micron$) is about 10\%,
substantiating the assumption that the corrugation of the electric
field close to the mirror surface can be neglected. 

\begin{figure}
\centering
\includegraphics[width=.45\textwidth]{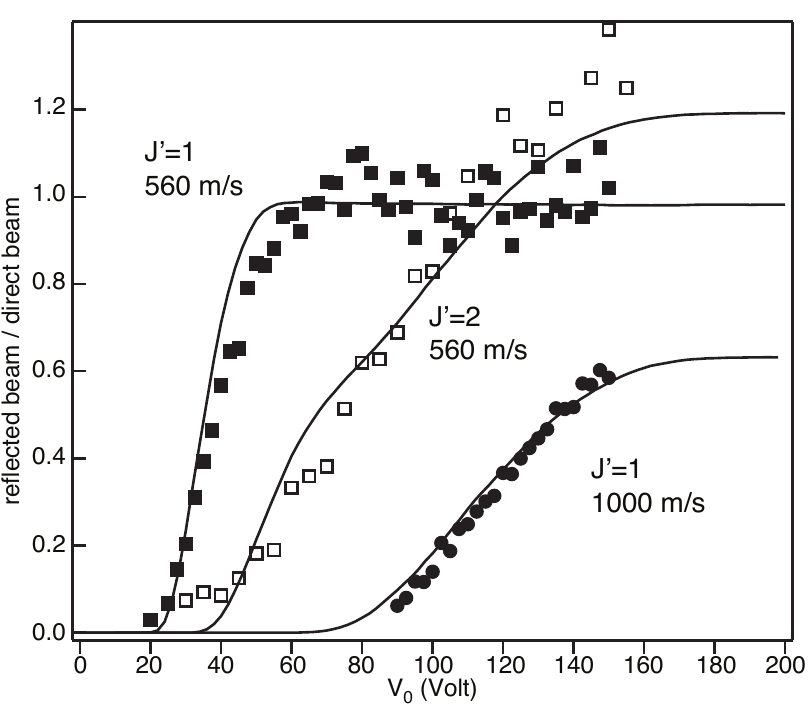}
\caption{Ratio of the signal of the reflected beam and the direct beam 
  as a function of the voltage $V_0$ applied to the electrodes for a 
  fixed position of the elliptical mirror, both from experiments (markers) 
  and simulations (lines). Filled circles are the data for a beam of metastable 
  CO molecules in the $J'=1$ level with a mean velocity of 1000~m/s. Filled 
  (empty) squares are the data for metastable CO molecules in the $J'=1$ 
  ($J'=2$) level with a mean velocity of 560~m/s. 
\label{fig:reflectivity}}
\end{figure}

In another set of measurements we have studied the reflectivity of the
mirror at a fixed position and orientation, for beams of metastable CO molecules in different 
rotational levels and with different mean velocities, as a function of the 
voltages applied to the electrodes. The position of the mirror was similar 
to the one that returned the data of Figure~\ref{fig:tilt} for $\chi=0^\circ$, 
where the groups of molecules that reach the detector directly 
and those that reach it only after being reflected are clearly separated.
In this position, the mean angle of incidence of the molecules on the mirror 
is between $2.4^\circ$ and $2.7^\circ$. 
The collected data (markers) are shown in 
Figure~\ref{fig:reflectivity} together with the outcome of trajectory
simulations (lines). The data-points represent the ratio of the number of
molecules in the reflected beam to that in the direct beam. The full (empty)
squares are the data for metastable CO molecules in the $J'=1$
($J'=2$) level seeded in krypton, and the circles are the data for a beam 
of metastable CO molecules in the $J'=1$ level seeded in neon. As 
expected, higher voltages are needed to reflect the faster CO molecules. 
The faster beam also yields lower overall ratios of the number of reflected
molecules to the number of molecules in the direct beam due to the
lower divergence. For a beam with a given mean velocity, lower
voltages are required to reflect CO molecules in the $J'=1$ level than
in the $J'=2$ level, in accordance with the larger effective dipole
moment of the former level (see Equation~\eqref{eq:mueff}). The upper
$\Lambda$-doublet component of the $J'=2$ level splits into four
low-field-seeking states (the doubly degenerate $|M'\Omega|=1$ and
$|M'\Omega|=2$ states) and one non-degenerate $M'=0$  state, whereas
the $J'=1$ level splits into two low-field-seeking states and one
non-degenerate $M'=0$ state. This explains the 20\% higher (4/5
\emph{versus} 2/3) asymptotic ratio in Figure~\ref{fig:reflectivity}
for the $J'=2$ level than for the $J'=1$ level. 

\section{Discussion and conclusion}
We have demonstrated that an electrostatic elliptical mirror can be
used to focus molecules coming through a circular skimmer opening into
a line-focus further downstream. The focusing properties of the mirror
and its dependence on the quantum state of the molecules, on the
velocity of the molecules and on the voltages applied to the
electrodes is quantitatively understood. Producing the elliptical
mirror with microstructured electrodes has the advantage that the
actual reflection takes place in a thin region above the surface, with
a thickness that is a fraction of the periodicity of the electrode
array. Near the outer edge of the reflective region, the corrugation
of the electric field strength can be neglected. When the mirror is
used under grazing incidence angles, as done here, the corrugation of
the electric field as experienced by the molecules is averaged out,
and can then also be neglected deeper in the reflective region, i.e.,
closer to the surface. The mirror therefore acts as a ``hard mirror''
and its effective reflective surface closely follows the shape of the
substrate on which the electrodes are positioned. We have used this
here to shape the originally flat structure into a plano-elliptical
structure, but other structures can certainly be made as well with
this approach. 

As stated in the Introduction, we plan to use this elliptical mirror
to increase the density of molecules trapped on a molecule chip. The
molecule chip we have used thus far is actually exactly the same as
the one used for the elliptical mirror, but instead of two  static
voltages ($\pm V_0$), six time-dependent (sinusoidal) voltages are then
applied to the electrodes. It has been detailed elsewhere that this
creates an array of 4~mm long, about 15--20\micron\ diameter
cylindrical traps for molecules in low-field-seeking quantum states
above the chip \cite{Meek:2008p153003,Meek:2009p055024}. Molecules from a
molecular beam can be confined in these tubular traps near the
entrance of the molecule chip, provided these traps move with the same
speed as the molecules in the molecular beam. When these traps are
subsequently decelerated to a standstill, the molecules stay confined
in the stationary traps above the chip \cite{Meek:2009p1699}. To
increase the density of molecules on the chip, we need to match the
spatial distribution of the molecular beam near the entrance of the
molecule chip to the tubular shape of the traps. It is clear from the
experiments that we have described here, that this is what the
elliptical mirror does. This alone is not sufficient, however, because
the focusing of the molecules in the $y$-direction necessarily leads
to an increase of the width of the $v_y$ velocity distribution in the
$y$-direction. What is needed is that the ($v_y$, $y$) phase-space
distribution near the entrance of the molecule chip matches the
corresponding phase-space acceptance of the tubular traps. 

\begin{figure}
\centering
\includegraphics[width=.45\textwidth]{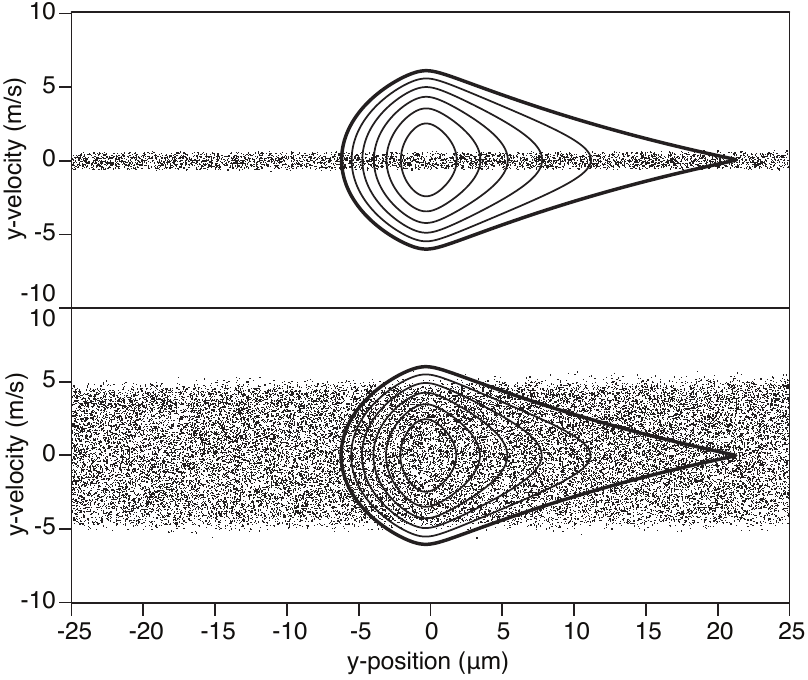}
\caption{Phase-space acceptance of the microtraps on the molecule chip
  in the direction perpendicular to the surface of the chip, i.e., the
  ($v_y$, $y$)-acceptance.  The dots in the upper (lower) panel
  indicate the calculated phase-space distribution when a
  beam of metastable CO molecules in the $J'=1$ level with a mean
  velocity of 300~m/s is coupled onto the chip, positioned 24.7~cm away
  from the 1~mm diameter source, without (with) the use of the
  elliptical mirror. 
  \label{fig:Phase-space}} 
\end{figure}

In Figure~\ref{fig:Phase-space} the calculated ($v_y$, $y$)
phase-space acceptance of the molecule chip is shown.
If the entrance of the molecule chip is positioned about 25~cm
away from the molecular beam source, as we have done thus far
\cite{Meek:2009p1699}, and the molecules are coupled onto the chip
without the elliptical mirror in place, the vertical phase-space
distribution of the beam is poorly matched to the vertical
acceptance. This is shown in the upper panel, where the width of the
$v_y$ velocity distribution (indicated by the dots) is much narrower
than what could be accepted by the traps. This changes dramatically
when the elliptical mirror is placed in between the beam source and
the molecule chip. As shown in the lower panel, the width of the $v_y$
velocity distribution now nicely matches the velocity acceptance of
the traps. The accompanying reduction of the spatial distribution in
the $y$-direction is irrelevant in this case, as the traps are
spatially overfilled anyway. Obviously, the molecule chip now needs to
be tilted relative to the molecular beam axis such that the
$v_y$-distribution is centered around zero; a tilt angle of about
3.2$^\circ$ is assumed in the simulations. It is seen from this
comparison that an increase in the density of molecules on the chip of
up to an order of magnitude can be expected when the elliptical mirror
is implemented. Apart from an improved matching of the beam to the
acceptance of the trap, the state selection provided by the mirror is
expected to be an important advantage for future experiments.\medskip

\begin{acknowledgments}
This work has been funded by the European Community's Seventh
Framework Program FP7/2007-2013 under grant agreement 216 774 and
ERC-2009-AdG under grant agreement 247142-MolChip. G.S. gratefully
acknowledges the support of the Alexander von Humboldt Foundation.  
\end{acknowledgments}

\bibliographystyle{gams-notit-nonumb}
\bibliography{./biblio}

\end{document}